\documentclass[aps,prd,floatfix,superscriptaddress,preprintnumbers,nofootinbib,preprint,tightenlines,longbibliography]{revtex4-1}
\pdfoutput=1

\usepackage{amsmath, amssymb, amsfonts, graphicx, mathrsfs, verbatim, float, slashed, bbm, color, xcolor, xspace, enumerate, url, bbding, multirow, outlines,upgreek}
\usepackage[english]{babel}
\usepackage[colorlinks=true,linkcolor=black,citecolor=darklightsabergreen]{hyperref}
\definecolor{darklightsabergreen}{rgb}{0.0, .49, 0.06}

\newcommand{\acro}[1]{\textsc{\MakeLowercase{#1}}} 
\newcommand{\osn}{\oldstylenums}

\newcommand{\gsim}{\gtrsim}
\newcommand{\lsim}{\lesssim}

\def\Oc{\mathcal{O}}

\newcommand{\beq}{\begin{equation}}
\newcommand{\eeq}{\end{equation}}
\newcommand{\bea}{\begin{eqnarray}}
\newcommand{\eea}{\end{eqnarray}}
\newcommand{\nn}{\nonumber}

\def\mdm{m_\chi}
\def\mDM{m_\chi}
\def\rhodm{\rho_\chi}
\def\vdm{v_{\rm \chi}}
\def\texp{t_{\rm exp}}

\def\Adet{A_{\rm det}}

\def\mtarget{m_{\rm T}}

\def\sigmaNDM{\sigma_{{\rm n} \chi}}
\def\sigmaT{\sigma_{{\rm T}\chi }}

\def\mDMmax{m_\chi^{\rm max}}

\def\muTDM{\mu_{{\rm T}\chi}}
\def\munDM{\mu_{{\rm n}\chi}}
\definecolor{darklightsabergreen}{rgb}{0.0, .49, 0.06}
\definecolor{orange}{rgb}{1.0, 0.5, 0.0}
\definecolor{galacticcenterbubblegum}{rgb}{.9, 0.2, 0.5}

\begin{document}

\title{Foraging for dark matter in large volume liquid scintillator neutrino detectors with multiscatter events}

\author{Joseph Bramante}
\affiliation{The McDonald Institute and Department of Physics, Engineering Physics,
and Astronomy, Queen's University, Kingston, Ontario, K7L 2S8, Canada}
\affiliation{Perimeter Institute for Theoretical Physics, Waterloo, ON N2J 2W9, Canada}

\author{Benjamin Broerman}
\affiliation{The McDonald Institute and Department of Physics, Engineering Physics,
and Astronomy, Queen's University, Kingston, Ontario, K7L 2S8, Canada}

\author{Jason Kumar}
\affiliation{Department of Physics and Astronomy, University of Hawaii, Honolulu, Hawaii 96822, USA}

\author{Rafael F.~Lang}
\affiliation{Department of Physics and Astronomy, Purdue University, West Lafayette, IN 47907, USA}

\author{Maxim Pospelov}
\affiliation{Perimeter Institute for Theoretical Physics, Waterloo, ON N2J 2W9, Canada}
\affiliation{Department of Physics and Astronomy, University of Victoria, Victoria, BC V8P 5C2, Canada}

\author{Nirmal Raj}
\affiliation{TRIUMF, 4004 Wesbrook Mall, Vancouver, BC V6T 2A3, Canada}

\begin{abstract}
We show that dark matter with a per-nucleon scattering cross section $\gtrsim 10^{-28}~{\rm cm^2}$ could be discovered by liquid scintillator neutrino detectors like \acro{borexino}, \acro{sno}+, and \acro{juno}. 
Due to the large dark matter fluxes admitted, these detectors could find dark matter with masses up to $10^{21}$~GeV, surpassing the mass sensitivity of current direct detection experiments (such as \acro{Xenon}\osn{1}\acro{T} and \acro{PICO}) by over two orders of magnitude.
We derive the spin-independent and spin-dependent cross section sensitivity of these detectors using existing selection triggers, and propose an improved trigger program that enhances this sensitivity by two orders of magnitude.
We interpret these sensitivities in terms of three dark matter scenarios: (1) effective contact operators for scattering, (2) \acro{QCD}-charged dark matter, and (3) a recently proposed model of Planck-mass baryon-charged dark matter. 
We calculate the flux attenuation of dark matter at these detectors due to the earth overburden, taking into account the earth's density profile and elemental composition, and nuclear spins.
\end{abstract}

\maketitle


\section{Introduction}
Despite ongoing experimental inquiry, the characteristics and couplings of dark matter remain a compelling puzzle. In this document we show how dark matter may be discovered by looking for multiple-scatter nuclear recoil events at large volume liquid scintillator experiments, which are traditionally used to detect solar neutrinos with MeV energies.

Ongoing dark matter direct detection experiments typically seek dark matter that scatters in underground detectors at most once. These are often called  \acro{WIMP}s, or weakly interacting massive particles, and their signature is a recoiling nucleus with $\sim {\rm keV}$ kinetic energy.
As some of us recently showed, traditional dark matter direct detection experiments are also capable of seeking dark matter candidates that scatter multiple times as they transit these detectors \cite{1803.08044}. 
Because they have a large cross section with Standard Model particles, these \acro{MIMPs}, or multiply interacting massive particles, leave a signature different  from \acro{WIMP}s. 
M\acro{IMP}s recoil against nuclei multiple times as they travel through dark matter direct detection experiments, leaving a characteristic collinear trail of deposited nuclear recoil energy.
M\acro{IMP} searches extend the reach of dark matter direct detection searches to the largest dark matter masses; the maximum dark matter mass an experiment can search for is determined by the total dark matter flux admitted by the detector, which in turn is determined by the area of the detector and the observation time.
Current xenon-based and bubble chamber experiments may eventually probe dark matter up to Planck masses ($M_{\rm Pl}$). 
Such dark matter candidates that are super-heavy and strongly interacting arise in supersymmetric theories \cite{Raby:1997pb}, are producible out of equilibrium in the early universe \cite{Chung:1998zb,Harigaya:2016vda,Kolb:2017jvz}, and as we will show in this paper, arise naturally in composite dark matter theories.

\begin{figure*}
\includegraphics[width=.7\textwidth]{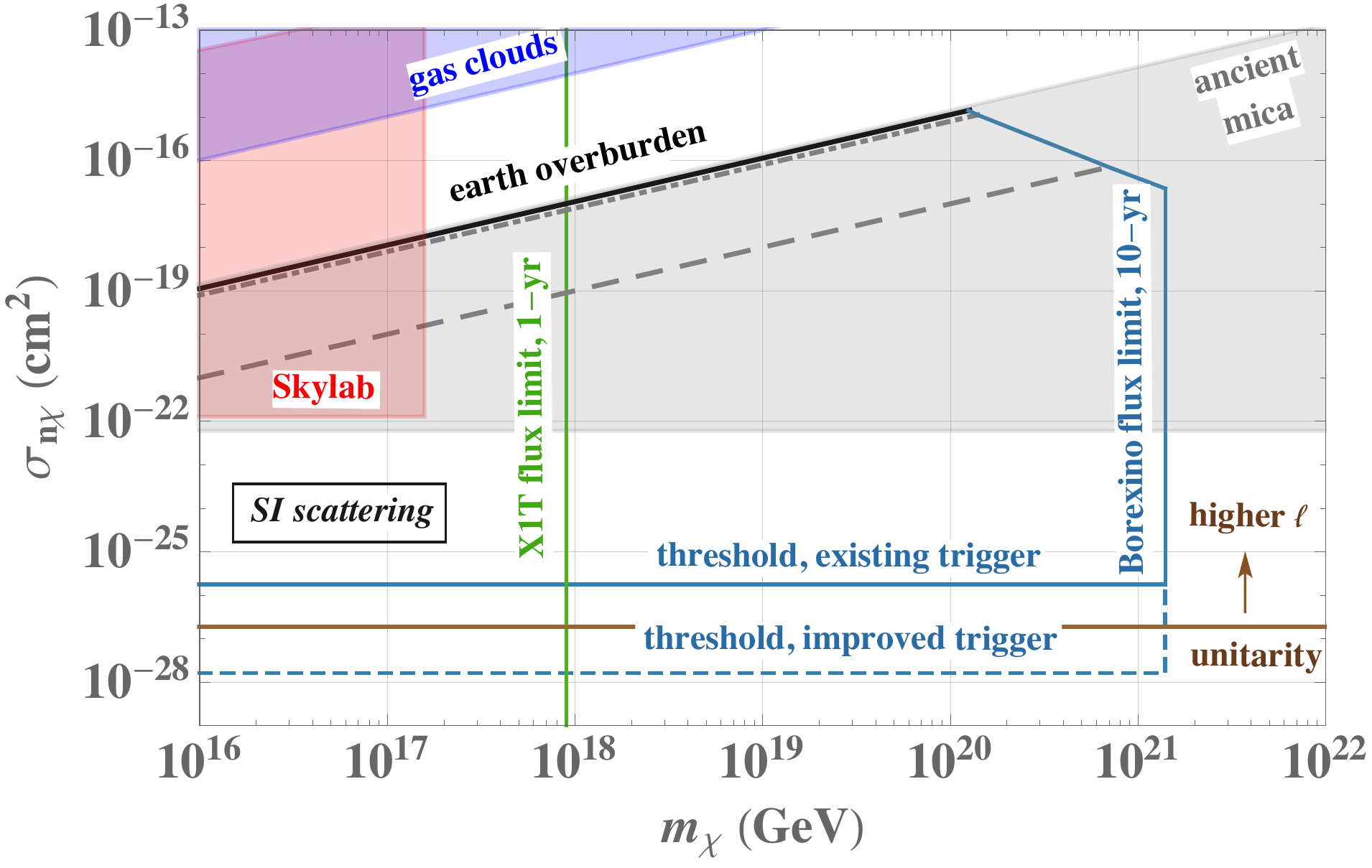} 
\includegraphics[width=.7\textwidth]{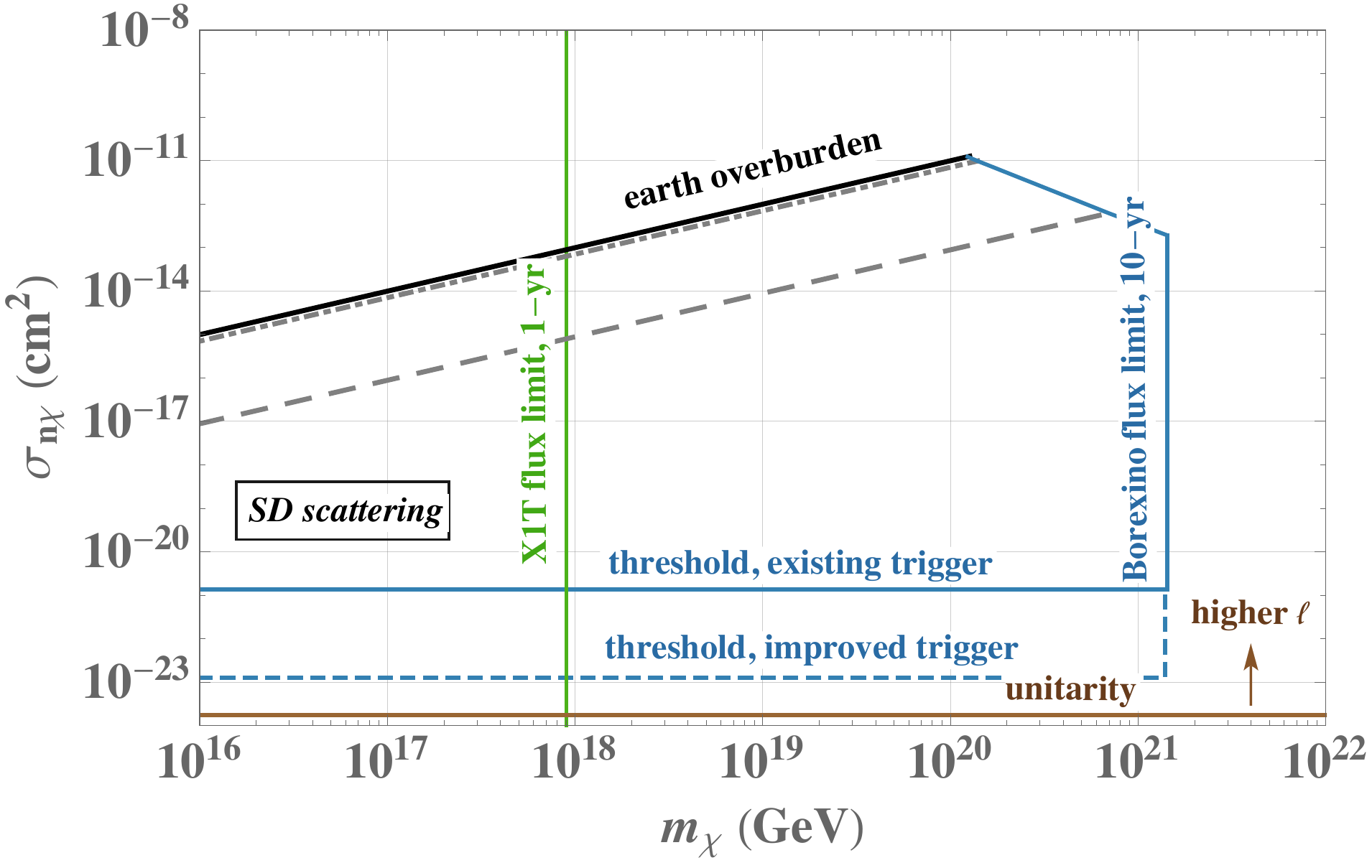}
\caption{Sensitivities, as derived in Sec.~\ref{sec:reaches}, of liquid scintillator detectors to multiscatter-inducing dark matter undergoing spin-independent (\textbf{\em top}) or spin-dependent (\textbf{\em bottom}) scattering, illustrated with \acro{borexino}.
The mass reach is determined by the total flux of dark matter transiting the detector, while the detection threshold cross section is set by the minimum rate of energy deposition required for the photomultiplier tubes to tell significantly between signal scintillation and dark counts.
Due to similar dimensions and detection technology, \acro{sno}+ would have very similar sensitivities; due to larger dimensions, \acro{juno} would reach greater dark matter masses.
The vertical green line is the maximum mass reach of \acro{Xenon}\osn{1}\acro{T} after 1 year \cite{1803.08044}.
The black lines indicate upper bounds (derived in Sec.~\ref{sec:ob}) on the cross section from the slowing down of dark matter by the earth's atmosphere and rock.
The dot-dashed (dashed) lines indicate the halving (tenthing) of total dark matter flux by this earth overburden.
Also shown are spin-independent scattering bounds from 
molecular gas cloud cooling \cite{1806.06857}, 
Skylab \cite{skylab,Starkman:1990nj,McGuire:1994pq}, and
searches for dark matter tracks in ancient mica \cite{Price:1986ky,1410.2236}. 
Similar bounds should apply to spin-dependent scattering; this would require a re-analysis of these constraints accounting for the fraction of spin-containing nuclei in gas clouds, Skylab detectors, and ancient mica.
The horizontal brown line denotes the maximum cross section imposed by unitarity at zero angular momentum; this cross section could be higher for higher $\ell$ modes.
}
\label{fig:reaches}
\end{figure*}

This paper will show that dark matter with mass {\em greater} than $M_{\rm Pl}$ can be readily sought in a number of existing and imminent large volume neutrino detectors, such as \acro{BOREXINO, sno+}, and \acro{juno}. 
While the energy threshold of neutrino detectors is typically larger than dedicated dark matter experiments, their larger volumes collect over three orders of magnitude more dark matter flux.
Of course, neutrino detectors are unsuitable for a standard \acro{WIMP} search, where a single $\Oc$(10 keV) recoil energy deposition falls well short of $\Oc$(MeV) thresholds that trigger these experiments. 
However, dark matter with sufficiently many scattering interactions will impart enough aggregate kinetic energy to trigger these detectors, rendering them well suited for \acro{MIMP} searches.
Their cross section vs mass reach is given by the region above and to the left of the blue lines in Fig.~\ref{fig:reaches}.
We will calculate this reach in the following.  

Other proposed searches for dark matter that deposits enough energy to exceed the $\sim$ MeV threshold of neutrino detectors include dark matter 
that destroys target baryons \cite{1008.2399,1106.4320,An:2012bs,1312.0011}, 
that yields annihilation or decay products detectable in these experiments \cite{Yuksel:2007ac,1405.7370,1410.2246,1411.6632,1611.09866,1711.04531,1711.05278,Campo:2017nwh,1804.07302,1812.05102,Berger:2018urf}, 
that deposits its entire (mass + kinetic) energy \cite{1712.00455},
is produced at high-intensity accelerators or radioactive sources \cite{1405.4864},
is accelerated by astrophysical sources \cite{1611.04599},
or is bounced off energetic cosmic rays \cite{1810.10543,Ema:2018bih}. 
All these approaches (excepting the last one) require specific models with a number of massive dark sector states. In contrast, our proposed search is model-independent and sensitive to any single dark matter state with a large nuclear scattering cross section over a range of masses.  
Another important difference is that in our study, \v{C}erenkov light-based detectors like \acro{SNO} and Super-Kamiokande will not be effective, since \acro{MIMP} scattering does not excite electrons/nuclei to relativistic speeds. Future argon-based detectors such as Proto\acro{dune} and \acro{dune}, if operated in single-phase mode with particle energies measured via electronic readout, may also be unsuitable to our study, since the energy threshold is around 5 MeV for even a single scatter. Lowering this threshold by operating in dual-phase may make these detectors more interesting for \acro{mimp} searches.

The remainder of this document proceeds as follows.  
In Sec.~\ref{sec:reaches}, we detail the dark matter-nucleon scattering cross sections that could be uncovered using large volume liquid scintillator detectors. We consider both a conservative search, which requires dark matter passing the triggers defined at these experiments, and an aggressive future search, where the irreducible background at large volume liquid scintillator experiments is estimated using the dark count rate in photomultiplier tubes (\acro{pmt}s) as reported by \acro{Borexino}. 
In Sec.~\ref{sec:models} we discuss three scenarios of super-heavy dark matter that scatters multiple times in and could be discovered by large volume liquid scintillator experiments. 
In Sec.~\ref{sec:ob} we provide overburden calculations, determining the fraction of multiply interacting dark matter that is blocked from detection by scattering with the earth.
In Sec.~\ref{sec:concs} we conclude.

\section{Liquid scintillator searches for dark matter}
\label{sec:reaches}
Large volume liquid scintillator experiments like \acro{Borexino}, \acro{SNO}+, and \acro{JUNO} can be used to detect strongly interacting, heavy dark matter. 
The dark matter scattering multiplicity must be large enough that enough energy is deposited to exceed the trigger threshold of these experiments, which are primarily designed to detect neutrinos with $\gsim$ MeV energies. The dark matter must be heavy enough that its initial kinetic energy greatly exceeds the energy lost scattering with the intervening atmosphere and earth as it travels to these experiments situated a few kilometers underground. 
The advantage of large liquid scintillator experiments is that they are a few orders of magnitude larger in dimension than traditional direct detection experiments, and therefore intercept (a few)$^2$ orders of magnitude larger flux of dark matter.

To explore sensitivities to dark matter at these experiments, we begin with the canonical, momentum-independent spin-independent and spin-dependent dark matter-nucleon scattering cross sections. 
In Sec.~\ref{sec:models} we discuss a number of explicit dark matter models which permit large masses and sizable nuclear scattering cross sections. For a target with atomic number $A$, dark matter-nucleus reduced mass $\muTDM$, and dark matter-nucleon reduced mass $\munDM$, the spin-independent \acro{SI} and spin-dependent \acro{SD} per-nucleon cross sections in terms of the per-nucleus cross section are given by
\bea
\nn \sigmaNDM^{({\rm \acro{SI}})} &=& \left(\frac{\munDM}{\muTDM}\right)^2  \left(\frac{1}{A^2}\right) \sigmaT \;\; \propto \;\; \left(\frac{1}{A^4}\right) \sigmaT~,\\
\nn \sigmaNDM^{({\rm \acro{SD}})} &=& \left(\frac{\munDM}{\muTDM}\right)^2  \left[\frac{4}{3} \frac{J_A + 1}{J_A} [\langle S_p \rangle + \frac{a_n}{a_p} \langle S_n \rangle]^2\right]^{-1} \sigmaT \\ 
&\propto& \left(\frac{1}{A^2}\right) \sigmaT~,
\label{eq:pernucleonXS}
\eea
where the proportionality is in the $\mDM \gg \mtarget$ limit, and $\langle S_p \rangle, \langle S_n \rangle$ are the nuclear matrix elements of nucleon spin.
The effect of nuclear form factors is negligible in our treatment, hence they have been ignored here. 
We will consider isospin-independent scattering and set $a_n=a_p$, although it would be possible to extend these results to the case $a_n \neq a_p$.
The above formula only applies to scenarios where perturbativity is valid; for non-perturbative scenarios, such as when dark matter is comprised of large bound states, only per-nuclear cross sections are relevant.

Because dark matter will only have to scatter (multiple times) to be detected at \acro{SNO}+, \acro{Borexino}, and \acro{JUNO}, a determination of which dark matter parameters can be probed by these detectors requires computing just two quantities: 
(i) the maximum dark matter mass to which the experiment is sensitive, and 
(ii) the minimum cross section for the dark matter to exceed the energy thresholds of the experiments.
We will estimate these two quantities in the following, assuming uniform dark matter velocity $\vdm = 10^{-3} c$. 
This is a good approximation to a full estimation of these quantities assuming a Maxwell-Boltzmann speed distribution, which we will encounter in Sec.~\ref{sec:ob}. 

\subsection{Mass reach}
\label{sec:mass}

The maximum dark matter mass that large liquid scintillator detectors are sensitive to, $\mDMmax$, depends upon the detectable dark matter flux passing through the detector. Because the dark matter density around earth is constant, the largest dark matter mass probed by the detector will correspond to a flux that delivers a few detectable dark matter events over the observation time of the detector,
$(\rhodm/\mdm) \vdm \Adet \texp \simeq$ 1, where 
$\rhodm$ = 0.3~GeV/cm$^3$~\cite{2018arXiv181009466E},
$\vdm$
and $\mdm$
are respectively the local density,
average velocity,
and mass of incident dark matter, 
$\Adet$ is the effective area of the fiducial volume of the detector, and 
$\texp$ is the observation time.
From this, we obtain 
\bea
\mDMmax = 1.4 \times 10^{21} \ {\rm GeV} \left(\frac{\Adet}{5 \times 10^5 \ {\rm cm}^2}\right) \left(\frac{\texp}{10 \ {\rm yr}}\right)~.
\eea
Here we have normalized  $\Adet$ to the area of the spherical, $8.5 ~{\rm meter}$ diameter, ``inner volume" region of  \acro{borexino} \cite{1308.0443}, which yields $\Adet = \pi r^2 \simeq 5 \times 10^5$ cm$^2$. 
We have indicated the \acro{borexino} 10-year sensitivity with a vertical blue line in Fig.~\ref{fig:reaches}. 
At \acro{sno}+, $\Adet \simeq 10^6$ cm$^2$, hence $\mDMmax \simeq 2.8 \times 10^{21}$~GeV after 10 years. 
Similarly, at \acro{juno}, $\Adet \simeq 10^7$ cm$^2$, hence $\mDMmax \simeq 2.8 \times 10^{22}$~GeV after 10 years.

For a sufficiently large dark matter-nucleon scattering cross section, the mass reach of large volume scintillator experiments will diminish, since a portion of the dark matter flux will become undetectably slow after scattering many times with nuclei in the earth. This diminished mass reach is also shown in Fig.~\ref{fig:reaches}. We discuss this further in Sec.~\ref{sec:ob}.

\subsection{Cross section reach with existing triggers}
\label{subsec:triggerexisting}

Next we determine the minimum dark matter-nucleon cross section required to trigger large volume liquid scintillatior neutrino detectors. 
At some of these detectors, this threshold is quoted in terms of the energy deposited per $\sim$100 ns window. 
At \acro{borexino}, the detector is triggered if $E_{\rm trig} \simeq$ 50 keV of energy is deposited within a 100 ns timing window \cite{1308.0443}. 
In this study, we consider a dark matter flux which travels $3-10$ meters through these detectors over a time $10-30$ $\mu$s. 
We note that assuming the dark matter travels at least 3 meters alters the flux calculations given above by less than 10\%. 
The signal of such a \acro{MIMP} would be a continuous emission of photons across a succession of many $\Oc(100)$ ns timing windows. 
The typical lab frame recoil energy of a carbon nucleus struck by heavy dark matter  $= m_{\rm T} \vdm^2$ in the limit $\mdm \gg m_{\rm T}$, where $m_{\rm T}$ is the mass of the target nucleus. The density of the scintillator pseudocumene (C$_9$ H$_{12}$) is 0.88 g/cm$^3$. Therefore, using the mean scattering rate formula for scattering on carbon with number density $n_{\rm C}$ ($\Gamma_s = n_{\rm C} \sigma_{\rm C \chi} \vdm$), the 50 keV per 100 ns trigger threshold amounts to a per-carbon-nucleus cross section of $\sigma_{\rm C \chi} = 3.7 \times 10^{-23}$ cm$^2$. 
From Eq.~\eqref{eq:pernucleonXS}, this corresponds to a per-nucleon scattering cross section
$
\sigmaNDM^{(\rm \acro{SI}, min,~ 100\%~eff. )} = 1.8 \times 10^{-27}~{\rm cm^2},
$
set by scattering on $^{12}$C.
The threshold for a spin-dependent cross section is set by scattering on $^{13}$C. Accounting for its natural abundance = 1.1\%,
and from Eq.~\eqref{eq:pernucleonXS}, this threshold is given by $\sigmaNDM^{(\rm \acro{SD}, min,~ 100\%~eff. )} = 1.4 \times 10^{-22}~{\rm cm^2}$.
We have taken the values $\{J_A, \langle S_p \rangle, \langle S_n \rangle\} = \{1/2, -0.026, -0.155\}$ for $^{13}$C \cite{Bednyakov:2004xq}.
The effect of \acro{sd} scattering on (the more abundant) $^1$H is relatively negligible, since the energy deposition is small, and since Eq.~\eqref{eq:pernucleonXS} implies poor sensitivity to small per-nucleon cross sections for $A$=1.
Note that this is because we had assumed that $a_p = a_n$ in Eq.~\eqref{eq:pernucleonXS}; were $a_n = 0$, the \acro{SD} scattering sensitivity would be set by $^1$H recoils.

However, these cross sections have assumed a 100\% efficiency for conversion of carbon nuclear recoil energy into photons detectable by \acro{PMT}s. 
In reality, we expect only about 10\% of the carbon recoil energy to be converted to detectable photons. 
In a study of liquid scintillation nuclear recoils \cite{Hong:2002ec}, the photoelectron yield for carbon and hydrogen nuclear recoils were determined for recoil energies as small as $40~{\rm keV}$. 
Extrapolating their result using their modified Birks' formula, the light yield efficiency for $10-20$ keV nuclear recoils is approximately 10\%.
This results in the scaling up of the minimum cross sections quoted above by a factor of ten.

Therefore, the \acro{Borexino} trigger outline above implies that heavy dark matter can be detected with a minimum per-nucleon spin-independent scattering cross section given by
\bea
\sigmaNDM^{(\rm \acro{SI}, min)} = 1.8 \times 10^{-26}~{\rm cm^2} \left(\frac{E_{\rm trig}}{50~{\rm keV}} \right), 
\label{eq:trig1SI}
\eea
and a minimum spin-dependent scattering cross section given by
\bea
\sigmaNDM^{(\rm \acro{SD}, min)} = 1.4 \times 10^{-21}~{\rm cm^2} \left(\frac{E_{\rm trig}}{50~{\rm keV}} \right), 
\label{eq:trig1SD}
\eea
where $E_{\rm trig}$ is the 100 ns trigger energy of the detector. 
We have plotted these threshold cross sections in Fig.~\ref{fig:reaches} with solid blue horizontal lines.
We have also shown existing constraints on spin-independent scattering from molecular gas cloud cooling \cite{1806.06857}, as well as from searches for dark matter tracks in plastic etch detectors on board Skylab \cite{skylab,Starkman:1990nj,McGuire:1994pq}, and in ancient muscovite mica \cite{Price:1986ky,Starkman:1990nj};
see Appendix~\ref{app:mica} for details of our recasting of mica bounds.
Also shown in Fig.~\ref{fig:reaches} is the maximum cross section dictated by unitarity, which at the per-nucleus level is $4\pi(2\ell+1)/q^2_{\rm C, com}$, for the partial wave mode $\ell =0$, with $q_{\rm C, com} = \mu_{{\rm C}\chi}\vdm$ the center-of-mass momentum of the dark matter-carbon nucleus system.
This cross section could of course be higher if scattering proceeds in higher partial wave modes.
We have displayed the per-nucleon unitarity cross sections after appropriately rescaling them using Eq.~\eqref{eq:pernucleonXS}.
The black and gray lines are cross sections above which the earth overburden significantly degrades dark matter kinetic energies; we will treat this more carefully in Sec.~\ref{sec:ob}.
We remark here that the (smaller-than-\acro{BOREXINO}) liquid scintillator sub-detectors of the erstwhile \acro{MACRO} experiment \cite{Ambrosio:2002qq} would have been sensitive to \acro{MIMP} parameter space; we reserve for future work the derivation of this sensitivity by carefully extracting the threshold and accounting for \acro{MACRO}'s non-trivial detector geometry.

Note that, the efficiencies obtained in \cite{Hong:2002ec} imply that on average one photoelectron will be emitted per $\sim 10$ keV carbon-atom recoil.
This means, with a 10\%-efficient light yield, the \acro{borexino} trigger is activated if a minimum of 50 photoelectrons are detected in a 100 ns window, or 5000 in a 10 $\mu$s window.
In the next subsection we argue that a photoelectron count that is $\Oc(100)$ smaller can be tolerated in \acro{mimp} searches, with the use of new trigger software.

\subsection{Cross section reach with improved triggers}
\label{subsec:triggeraggro}

It should be possible to improve the cross section sensitivity of large volume liquid scintillator experiments with the addition of a dedicated trigger. 
As long as this trigger is sensitive to excessive photoelectrons detected within a certain time window, \acro{mimp}s may be sought. 
As discussed, the window of time for a dark matter particle to transit these detectors is $\sim$ 10 ${\rm \mu s}$. 
The \acro{PMT} dark count rate reported by \acro{Borexino} \cite{1308.0443} is roughly 10 photoelectrons per 10 $\mu$s. 
Using this, one may now determine the minimum cross section -- or minimum multiplicity of scatters -- potentially accessible at a similar future experiment, such as \acro{SNO+}.
Assuming 10 $\mu$s timing windows, we will demand that the aggressive trigger threshold admit no more than a few randomly generated dark count events over the course of the experiment.
Said differently, we demand that the dark matter cross section sought with the aggressive trigger be just large enough that the maximum dark counts in any 10 $\mu$s window over the lifetime of the experiment, will only produce about one background event.
Then, assuming Poisson statistics the required multiplicity $N_{\rm c}$ is obtained by solving
\beq
\sum_{N_{\rm c}}^\infty \frac{(N_{\rm bg})^{N_{\rm c}}}{N_{\rm c}!} e^{-N_{\rm bg}} =  \frac{10~\mu{\rm s}}{t_{\rm life}}~,
\eeq
where $N_{\rm bg} =10$ is the dark count in a 10 $\mu$s window as mentioned above, and $t_{\rm life}$ is the run-time of the experiment.
For $t_{\rm life}$ = 10 years, we get $N_{\rm c} = 42$, which corresponds to a minimum detectable spin-independent cross section of
\bea
\sigmaNDM^{(\rm \acro{SI} \ min, \ trig+ )} \simeq 1.7 \times 10^{-28}~{\rm cm^2}~, \label{eq:trig2SI}
\eea
and a minimum detectable spin-dependent cross section of
\bea
\sigmaNDM^{(\rm \acro{SD} \ min, \ trig+ )} \simeq 1.3 \times 10^{-23}~{\rm cm^2}~. \label{eq:trig2SD}
\eea
We have plotted these improved threshold cross sections in Fig.~\ref{fig:reaches} with dashed blue horizontal lines.

\textbf{{\em $^{14}$C background.}} The $\beta$ decay of carbon-14, which produces $\sim 20-150$ keV electrons, has the highest rate of background processes observed at underground liquid scintillator experiments (see $e.g.$ Section XI.3 in \cite{1308.0443}). 
Using liquid scintillator refined from petroleum extracted deep underground, an abundance of $^{14}$C/$^{12}$C as low as $2 \times 10^{-18}$ has been obtained at \acro{Borexino}~\cite{Alimonti:1998rc}. 
With this purity, a kiloton of liquid scintillator will see $^{14}$C decays at a rate of about one kHz. 
Additionally, we note that \acro{PMT} hits from $\beta$ decays last $< \sim$100 ns~\cite{1308.0443}. 
Therefore, we should expect a $^{14}$C $\beta$ decay event in around one out of every hundred 10-$\mu$s \acro{MIMP} search windows. 
However, we should not expect more than 4 $^{14}$C events in any 10 $\mu$s window over the lifetime of the detector, meaning that we expect at least $10 - (4\times0.01)$ = 9.6 $\mu$s of every 10 $\mu$s window to be background-free (except for dark counts, which we have discussed above).  
Using this, and the fact that the observed energy deposition rate for $^{14}$C $\beta$ decays is a few orders of magnitude higher than the energy deposition for a \acro{MIMP} in our aggressive search region ($\sigmaNDM^{(\rm \acro{SI} \ min, \ trig+ )} \lesssim 10^{-26}~{\rm cm^2}$), it should be straightforward to design a suitable trigger for the aggressive \acro{MIMP} search region that rejects 10 $\mu$s windows with any $^{14}$C $\beta$ decays. 
In principle, it should also be possible to design the aggressive trigger to ``scrub" $^{14}$C $\beta$ decays, by excluding 100 ns windows with \acro{PMT} hits occurring in excess of a certain threshold. 
The particulars of the aggressive trigger will depend on the experimental capabilities of $e.g.$  \acro{Borexino} , \acro{SNO+}, \acro{JUNO}, and we leave a detailed study to future work.

\subsection{Directionality of recoiling nuclei in MIMP searches}
\label{subsec:direc}

Multiply interacting dark matter, with a per-nucleon cross section above the threshold cross sections identified in Eqs.~\eqref{eq:trig1SI}, \eqref{eq:trig1SD}, \eqref{eq:trig2SI} and \eqref{eq:trig2SD}, will scatter at least $40$ times, yielding a minimum of $40$ detectable photoelectrons (see discussion in Sec.~\ref{subsec:triggerexisting}), as it transits a large volume liquid scintillator-filled neutrino detector. 
The sites of these nuclear recoil interactions will lie along an extremely straight line pointing through the detector, since the deflection of a heavy particle by a carbon nucleus is, in the small-angle scattering limit,
\bea
\Delta \theta \lesssim \frac{m_{\rm T}}{m_{\rm \chi}} \simeq  10^{-16} \left( \frac{10^{17}~{\rm GeV}}{m_{\chi}} \right)  \left( \frac{m_{\rm T}}{11~{\rm GeV}} \right).
\eea

This collinearity of \acro{MIMP} events can be used to validate candidate super-heavy dark matter events, by using the location and timing of \acro{PMT} hits, and thereby reconstructing the track of the \acro{MIMP} as it transits the detector. While it is not feasible to resolve the directionality of the photons themselves, it has been demonstrated that by collecting the arrival time and \acro{PMT} location of all photon hits, directional information can be extracted; see $e.g.$ \cite{2016PhDT.......210S}, which showed how particle tracks in liquid scintillator can be reconstructed for the purposes of identifying neutrinos from dark matter that annihilates in the sun.
The precise details of such a directional analysis will depend on the data acquisition and timing capabilities of a given experiment.

\section{Survey of Multiply Interacting Dark Matter Models}
\label{sec:models}

There are a number of theories of dark matter that may be discovered at large volume liquid scintillator detectors. 
We will outline three scenarios here: an effective field theory description of dark matter, quantum chromodynamic \acro{(QCD)}-charged dark matter, and heavy dark matter mediated by a light vector carrying baryon number. 
A number of works demonstrate the production of super-heavy dark matter in the early universe~\cite{Kuzmin:1998kk,Chung:1998zb,Harigaya:2014waa,Kolb:2017jvz}.

\subsection{Effective operators}
\label{subsec:effops}
In Fig.~\ref{fig:reaches} and Eqs.~\eqref{eq:trig1SI},~\eqref{eq:trig1SD},~\eqref{eq:trig2SI} and~\eqref{eq:trig2SD}, the minimum cross sections required to trigger these detectors are a few orders of magnitude below those corresponding to \acro{qcd} hadronic scattering. 
We should therefore expect that a mediator transmitting these dark matter interactions has a minimum mass around the \acro{QCD} scale $\simeq {\rm 100~ MeV}$.

This intuition is borne out by the effective field theory cutoff for dark matter scattering interactions that just exceed the thresholds of large volume liquid scintillator experiments. 
Taking dark matter as a Majorana fermion $\chi$, we may characterize with effective operators its parton-level interactions with nucleons. 
For illustration we pick two operators: the scalar-scalar ($ss$) and axial-axial ($aa$) operators, describing respectively \acro{si} and \acro{sd} scattering in the non-relativistic limit: 
\bea
\nn \mathcal{O}_{ss} &=&  y_q \frac{(\bar{q}q)(\bar{\chi}\chi )}{\Lambda_{ss}^2}~, \\
\nn \mathcal{O}_{aa} &=& \frac{(\bar{q}\gamma_5\gamma_\mu q)(\bar{\chi}\gamma_5\gamma^\mu\chi)}{\Lambda_{aa}^2}~.
\eea
Following the usual convention (see \cite{Rajaraman:2011wf,1707.09442}), in order to evade stringent constraints from flavor observables, we assume the principle of minimal flavor violation \cite{Buras:2000dm}, whereby the $U(3)^5$ flavor symmetry of the Standard Model is broken only by Higgs Yukawa spurions.
Consequently, the scalar bilinears for quarks are proportional to the corresponding Higgs Yukawa coupling (or quark mass), and the axial bilinears are flavor universal.
 
We can now recast the threshold cross sections derived in Sec.~\ref{sec:reaches}, specifically assuming the new triggers proposed in Eqs.~\eqref{eq:trig2SI} and \eqref{eq:trig2SD}. 
The per-nucleon cross section for the above operators is given by \cite{Rajaraman:2011wf,1707.09442}
\bea
\nn \sigmaNDM^{ss} &=& \frac{4}{\pi} \munDM^2 \frac{(Z f^{ss}_p + (A-Z)f^{ss}_n)^2}{A^2}~,\\
\sigmaNDM^{aa} &=& \frac{3}{\pi} \munDM^2 |f^{aa}_n|^2~,
\eea
where the effective nucleonic couplings are 
\bea
\nn f^{ss}_{(n)} &=& \frac{\sqrt{2}m_n}{v \Lambda_{ss}^2} \left( \sum_{q = u, d, s} f^{(n)}_{T_q} + \sum_{Q=c,b,t} \frac{2}{27} f^{(n)}_{T_G}\right)~, \\
 f^{aa}_n &=& \frac{1}{\Lambda_{aa}^2}  \left( \sum_{q = u, d, s} \Delta^{(n)}_q \right)~,
\eea
with $v$ = 246 GeV the Higgs expectation value, and the nucleon form factors at zero momentum are taken as~\cite{Chang:2013oia,Dienes:2013xya} 
\bea
\nn \{f^p_{T_u},f^p_{T_d},f^p_{T_s},f^p_{T_G}\} &=& \{0.023,0.032,0.020,0.925\} ,\\
\nn \{f^n_{T_u},f^n_{T_d},f^n_{T_s},f^n_{T_G}\} &=& \{0.017,0.041,0.020,0.922\},\\
\nn \{ \Delta^n_u, \Delta^n_d, \Delta^n_s \} &=&  \{0.79, -0.32, -0.04\}.
\eea
Using this, we find that large volume liquid scintillator detectors like \acro{Borexino} and \acro{SNO}+ are potentially sensitive to the cutoffs
\bea
\nn \Lambda_{ss} &\leq& 59.8 \ {\rm MeV}~, \\
\Lambda_{aa} &\leq& 55.9 \ {\rm MeV}~.
\eea
  
Note that for the effective operator treatment to be valid, $\Lambda \gsim$ 11 MeV, the typical momentum transfer for dark matter scattering scattering off carbon. 

\subsection{QCD-charged dark matter}

The low-energy nucleon-on-nucleon elastic scattering cross section in the Standard Model ($\sim 10^{-26}~{\rm cm^2}$) is similar to the dark matter cross section sensitivity thresholds found for large volume liquid scintillator experiments in Sec.~\ref{sec:reaches}. 
Indeed, upon further investigation, dark matter charged under the Standard Model \acro{QCD} gauge group $SU(3)_{\rm C}$ provides a generic and viable class of models that can be sought out by these experiments.

Dark matter charged under $SU(3)_{\rm C}$ has been considered by a number of authors, $e.g.$ \cite{Raby:1997pb,Kang:2006yd,Jacoby:2007nw,DeLuca:2018mzn}. 
We will find that the super-massive ($\gtrsim 10^{17}$ GeV) dark matter considered here will likely need to transform as an octet of $SU(3)_{\rm C}$, once constraints on electromagnetically (\acro{EM}) charged particles from the \acro{MACRO} experiment \cite{Ambrosio:2000nq} are taken into account. 
The logic is as follows. 
Consider two color-charged dark matter states, a stable electroweak-singlet quark-like $X_3$, and a stable gluon-like $X_8$. 
Formally, these states transform as a triplet ($X_3$) and an octet ($X_8$) of $SU(3)_{\rm C}$. 
After the \acro{QCD} phase transition, $X_3$ will form color-neutral states such as $X_3 \bar q$ and $X_3 q q$, where $q$ are Standard Model quarks. Notice that while these $X_3$ composite states are color-neutral, they carry $\mathcal{O}(1)$ \acro{em} charge. 
The \acro{MACRO} experiment, by not observing minimum-ionizing particles in liquid scintillator tanks in the Gran Sasso mine, set a flux limit on heavy \acro{em} charged relic particles (see Fig.~1 of \cite{Ambrosio:2000nq}):
\bea
\Phi \lesssim 4 \times 10^{-15}~{\rm cm^{-2}~s^{-1}~sr^{-1}}~. \label{eq:macro}
\eea
Assuming that this search was sensitive to $\mathcal{O}(1)$ \acro{EM}-charged particles (see \cite{Perl:2001xi}), the \acro{MACRO} search excludes such particles with masses $10^8 ~{\rm GeV} \lesssim m_{X_3} \lesssim 10^{20}~{\rm GeV}$. 
This mass bound is an order of magnitude greater than quoted in \cite{Perl:2001xi}, possibly due to a transcription error in \cite{Perl:2001xi} that quotes the flux limit an order of magnitude higher than Eq.~\eqref{eq:macro} when deriving a bound from \acro{MACRO}. 

While we expect the lightest $X_3$-containing states ($X_3 \bar q$ and $X_3 q q$) to be \acro{EM}-charged, we expect the lightest $X_8$-containing states to be \acro{EM}-neutral. 
These states are $X_8 q \bar q $ and $X_8 g$, where $q$ is an up or down quark and $g$ is the Standard Model gluon. 
It is reasonable to assume that the lightest $X_8$-containing states are \acro{EM}-neutral, as is the case for the lightest pion in \acro{QCD}, and the $X_8 g$ state, which could be even lighter, is manifestly \acro{EM}-neutral. Detailed lattice computations will be necessary to establish the exact mass spectrum for $X_8 q \bar q $ and $X_8 g$ states.
The elastic scattering cross section for  $X_8 q \bar q $ and $X_8 g$, which will be surrounded by a cloud of quarks and gluons after confinement, should fall in the range
\bea
\sigma_{X_8} \sim 10^{-24} - 10^{-27}~{\rm cm^2}~,
\eea
where this is the typical range for nucleon-nucleon cross sections at low energy.
We defer to future work a complete cosmological treatment of superheavy \acro{QCD}-charged dark matter, as well as a lattice calculation to determine the  structure and mass spectrum of $X_8 q \bar q $ and $X_8 g$ bound states. 

\subsection{Baryo-charged dark matter with a sub-GeV mass mediator}

The effective operator treatment in Sec.~\ref{subsec:effops}, which found operator cutoffs of $\sim 100$ MeV for the cross section thresholds quoted in Sec.~\ref{sec:reaches}, indicates that dark matter scattering with $\lsim$ GeV mediators may be detected at large volume liquid scintillator experiments. 
Recently, Ref.~\cite{1809.07768} discussed models of Planck-mass \acro{mimp}s mediated by a light vector boson that carries gauged baryon number $U(1)_{\rm B}$, under which dark matter is charged. While such a light baryon-charged mediator is subject to a number of phenomenological constraints \cite{1809.07768,Dobrescu:2014fca,Jacques:2016dqz,Dror:2017ehi,Ismail:2017ulg,Dror:2017nsg}, they nevertheless allow for \acro{mimp}s that could be detected by large volume liquid scintillator experiments.
Assuming dark matter carries baryon charge $Q_\chi$,
for a $U(1)_{\rm B}$ gauge boson with coupling strength $\alpha_d$ 
and mass $m_{A_d}$, and normalizing with parameters unconstrained by meson decays,
the \acro{SI} per-nucleon scattering cross section is 
 \beq
 \sigma_{\rm baryo}^{\acro{SI}} \simeq 1.7 \times 10^{-28} \ {\rm cm}^2  
 \left(\frac{Q_\chi}{3}\right)^2 
 \left(\frac{\alpha_d}{0.01}\right)^2 
 \left(\frac{550 \ {\rm MeV}}{m_{A_d}}\right)^4~,
 \eeq
showing that \acro{borexino} and \acro{sno}+ are capable of probing this scenario.

\section{Earth Overburden Flux Attenuation}
\label{sec:ob}

\begin{figure*}
\includegraphics[width=.42\textwidth]{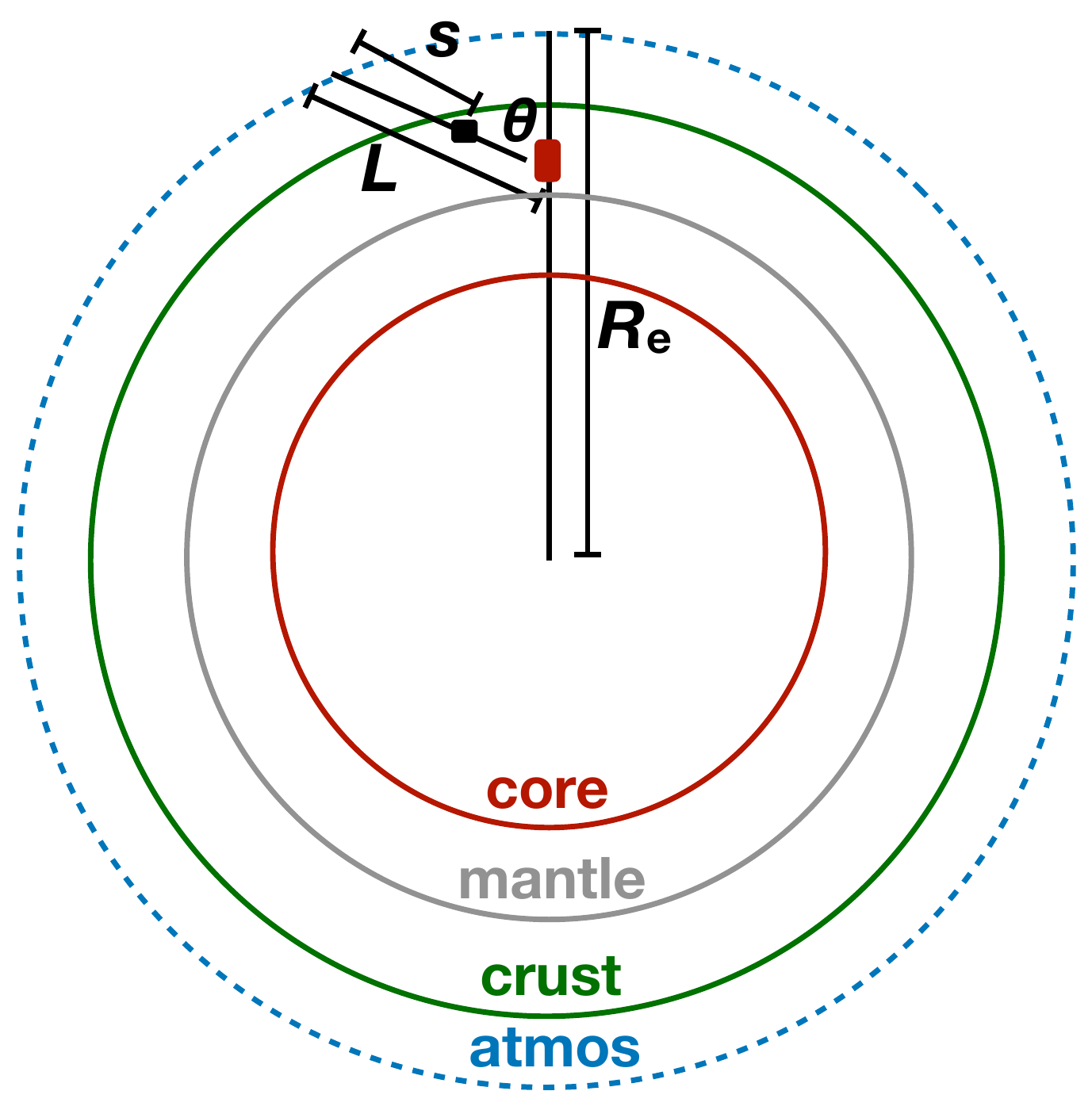}  \quad \quad
\includegraphics[width=.42\textwidth]{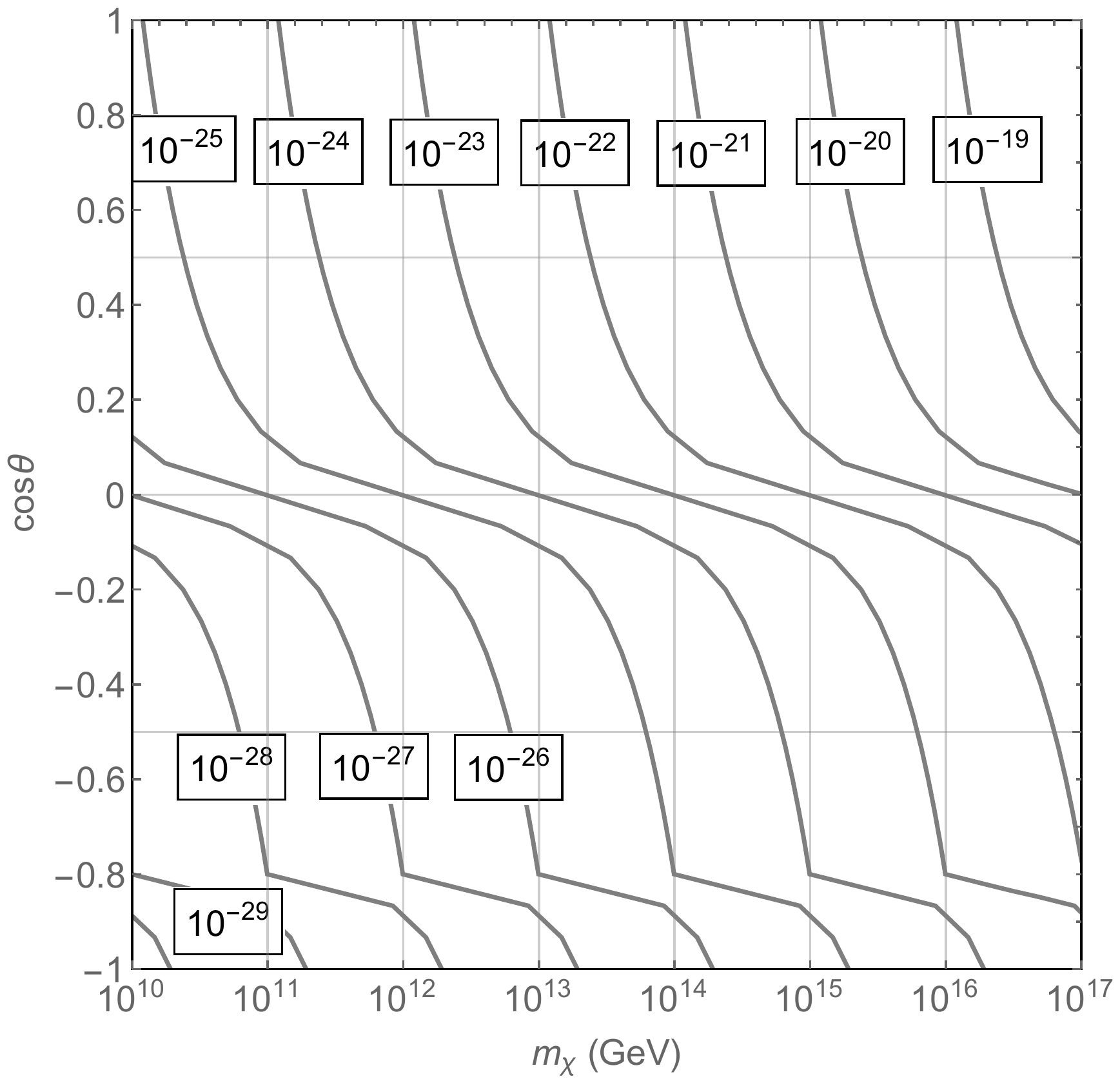}
\caption{
\textbf{\em Left.} Geometric schematic of a dark matter particle traveling along a straight path to a dark matter detector through the Earth's atmosphere and rock.
This diagram is not to scale.
\textbf{\em Right.} Contours in the $\cos\theta$-$\mdm$ plane, of the spin-independent per-nucleon cross sections (in cm$^2$) required for the Earth overburden to slow down dark matter below detector thresholds. 
This illustrative calculation assumes an isotropic dark matter distribution with a uniform speed = 220~km/s.}
\label{fig:ediagram}
\end{figure*}

If dark matter's cross section for scattering on Standard Model particles is sufficiently large, then dark matter particles will be slowed by repeated scattering while passing through the earth's atmosphere and crust.
For non-relativistic scattering, the fraction of kinetic energy depleted per scatter is $ z \beta \equiv 4 z (m_T \mdm)/(m_T + \mdm)^2$, where $z \in [0,1]$ is a kinematic factor that depends on the scattering angle (and $\mdm$, $\mtarget$ are respectively the dark matter and target masses).

Now consider a dark matter particle traversing a straight line path of length $L$ through the earth atmosphere and rock on its way to a detector,
\begin{align}
L = -r_{\rm det} \cos\theta + \sqrt{R_{\rm e}^2 - r_{\rm det}^2 \sin^2\theta}~,
\end{align}
where $r_{\rm det}$ is the distance between the detector and the center of the earth, $R_e$ is the radius of the earth (including its atmosphere), and $\theta$ is the angle between the incoming particle velocity and the line from the center of the earth to the detector; see the left-hand side of Figure \ref{fig:ediagram}.
At some distance $s$ along the path of length $L$ that the dark matter is traversing, the distance to the center of the earth $r$ is given by
\begin{align}
r^2 = R^2_{\rm e} + s^2 - 2 s (r_{\rm det} \cos\theta + L)~.
\label{eq:anglepath}
\end{align}

After traversing a path $L$, the final kinetic energy of the dark matter is given by
\begin{align}
E_f = E_i \prod_A \left(1-z \beta \right)^{\int_0^L n_A(r) \sigma_{A \chi} ds}~,
\label{eq:EfEi}
\end{align}
where $n_A(r)$ is the number density of nuclei of type $A$ at distance $r$ from the center of the Earth and we take an average value of $z = \frac{1}{2}$ in computations. 
The Earth and its atmosphere are assumed to be spherically symmetric, and our model of the density of Earth's interior follows the Preliminary Reference Earth Model \cite{1980PNAS...77.6973M}.
See also \cite{1803.08044} for the dimensions, densities and composition of the core, mantle and crust.
The $\langle S_p \rangle, \langle S_n \rangle$ values of spin-carrying elements are taken from \cite{Bednyakov:2004xq}.

 A good picture of the effects of the overburden may be gained by assuming an isotropic dark matter distribution with a uniform speed of 220 km/s.
 The length of dark matter's path through the Earth to the detector, related to $\cos\theta$ by Eq.~\eqref{eq:anglepath}, determines the cross section $\sigmaNDM$ above which $E_f$ is too low to trigger the detector.
For concreteness, we set the detector energy threshold to 1 keV.
 In the right hand side of Fig.~\ref{fig:ediagram} we show, in the $\cos\theta$-$\mdm$ plane, contours of $\sigmaNDM/{\rm cm}^2$ required for such a slow-down via spin-independent scatters.
 The visible change of slope near $\cos\theta = 0.1$ and $-0.8$ is due to the transit of dark matter through the mantle and core, which are (progressively) denser than the crust.
The overburden cross sections at $\cos\theta=1$ and $\cos\theta=-1$ are seen to agree with those derived in \cite{1803.08044}.
 For overburden cross sections between $\cos\theta=1$ and $\cos\theta=-1$, reconstructing directionality of dark matter in the detector (as discussed in Sec.~\ref{subsec:direc}) would provide a handle on the dark matter mass and cross section; see \cite{1803.08044} for more details.
 Above $\cos\theta = 0.2$ the slope of the contours is steep, indicating that the detectable dark matter flux becomes exponentially sensitive to the crustal overburden cross section.

 A more detailed calculation of the overburden bounds must take into account the dark matter velocity distribution and statistical considerations.
 To this end, we generated a Monte Carlo sample of dark matter particles 
 assuming an isotropic\footnote{Accounting for the directionality of dark matter affects our results at only the $\sim$10\% level, which would be invisible on a log-log plot of the nature of Figure~\ref{fig:reaches}.} population that follows the Maxwell-Boltzmann speed distribution (as seen by an observer moving through the halo)~\cite{Gould:1987ir}:
\begin{equation}
f(v) \propto   v \exp\left(\frac{-(v^2+v^2_\odot)}{v^2_{\rm disp}}\right)\left[\exp\left(\frac{2vv_\odot}{v^2_{\rm disp}}\right) - \exp\left(c_{\rm min}\frac{2vv_\odot}{v^2_{\rm disp}}\right) \right] \Theta(v_{\rm esc} + v_\odot - v)~,
\label{eq:MaxwellBoltzmann}
\end{equation}
where $v_\odot$ = 220 km/s is the speed of the sun through the halo,  
the velocity dispersion $v_{\rm disp} = 220$ km/s, and 
the galactic escape velocity $v_{\rm esc}$ = 533 km/s \cite{Piffl:2013mla}, and 
$c_{\rm min} = \max[-1,(v^2-(v^2_{\rm esc}-v_\odot^2))/2vv_\odot]$; $\Theta$ denotes the Heaviside theta function.
Bounds were obtained by locating ($\mDM$, $\sigmaNDM$) pairs that yield 2.3 expected \acro{MIMP}s over 10 years, with enough kinetic energy to deposit an average of 11 keV (the typical energy transfer to a carbon nucleus) in each carbon nuclear recoil as they transit \acro{Borexino}.
These are presented in Fig.~\ref{fig:reaches} as solid black lines. 
Due to the exponential sensitivity of the detectable flux to the overburden cross section (Eq.~\eqref{eq:EfEi}), the relation between this cross section and dark matter mass is practically linear, and can be approximated by
\bea
\nn \sigmaNDM^{\rm SI, ob} &\leq& 1.1 \times 10^{-19}~{\rm cm}^2\left(\frac{\mdm}{10^{16}~{\rm GeV}}\right)~,\\
\sigmaNDM^{\rm SD, ob} &\leq& 9.8 \times 10^{-16}~{\rm cm}^2\left(\frac{\mdm}{10^{16}~{\rm GeV}}\right)~.
\eea
This approximate linearity in Fig.~\ref{fig:reaches} is seen to break down near $\mdm = 10^{20}$~GeV; beyond this point, the dark matter flux is small, hence the blue line changes slope and tracks cross sections for which 2.3 \acro{mimp}s are detectable, until it joins the vertical blue line that denotes the dark matter mass for which the total flux is 2.3 \acro{mimp}s.
The dot-dashed and dashed gray lines denote, respectively, the cross sections for which the earth overburden reduces the detectable flux by a factor of 10 and 2.
These are approximated by
\bea
\nn \sigmaNDM^{\rm SI,\Phi/10} &=& 7.9 \times 10^{-20}~{\rm cm}^2\left(\frac{\mdm}{10^{16}~{\rm GeV}}\right)~,\\
\nn \sigmaNDM^{\rm SD, \Phi/10} &=& 6.9 \times 10^{-16}~{\rm cm}^2\left(\frac{\mdm}{10^{16}~{\rm GeV}}\right)~,\\
\nn \sigmaNDM^{\rm SI,\Phi/2} &=& 10^{-21}~{\rm cm}^2\left(\frac{\mdm}{10^{16}~{\rm GeV}}\right),\\
 \sigmaNDM^{\rm SD, \Phi/2} &=& 8.7 \times 10^{-18}~{\rm cm}^2\left(\frac{\mdm}{10^{16}~{\rm GeV}}\right)~.
\eea

One sees that $\sigmaNDM^{\rm SI,\Phi/2}$ corresponds to the cross section contours near $\cos\theta$ = 0 in the right hand side of Fig.~\ref{fig:ediagram}.
In presenting these overburden bounds, we have neglected the effect of ``saturated overburden scattering" \cite{1803.08044}, i.e. the insufficient slowing down of dark matter due to the finiteness of the number of terrestrial nuclei on its path to the detector.
This effect applies only to short-range interactions, whereas our treatment here is more generic; when this effect does apply, cross sections above the overburden lines in Fig.~\ref{fig:reaches} may also be probed by liquid scintillator neutrino detectors.

\section{Conclusions and Discussion}
\label{sec:concs}
It is sometimes the case that well-designed experiments are well-suited for novel physics searches beyond their intended scope.  
This document has shown that large volume liquid scintillator neutrino detectors stand at the vanguard of super-heavy dark matter detection. 
Because they are the largest terrestrial detectors capable of identifying a handful of nuclear recoil interactions over ambient backgrounds, experiments like \acro{borexino}, \acro{sno}+, and \acro{JUNO} provide unmatched high mass dark matter sensitivity. 
After estimating the sensitivity of these experiments to high mass dark matter in Sec.~\ref{sec:reaches}, we proceeded in Sec.~\ref{sec:models} to explore how canonical and newly proposed models of dark matter would first be discovered by large volume liquid scintillator experiments. 

We have also identified a major trigger improvement for future large volume liquid scintillator experiments, enabling the detection of dark matter-nucleon scattering cross sections over two orders of magnitude smaller than would be accessible with current triggers. 
We have shown that by recording experimental data whenever there are more than $40$ \acro{PMT} hits distributed over a 10 microsecond window, where this number of \acro{PMT} hits should never arise from randomly distributed \acro{PMT} misfires (``dark counts"), future liquid scintillator experiments can greatly increase their sensitivity to heavy dark matter.

It may be possible to further extend the sensitivity of large volume liquid scintillator experiments to even smaller nucleon-scattering cross sections, by incorporating directional information in these detectors. 
In Sec.~\ref{subsec:direc}, we demonstrated that multiply interacting dark matter traversing large volume liquid scintillator detectors will leave a characteristic, highly collinear trail of nuclear recoil sites, which will produce a pattern of scintillated photoelectrons depending on the timing and \acro{PMT} positions. 
Prior work on reconstructing the directionality of neutrinos from dark matter annihilating in the sun, using \acro{PMT}-instrumented liquid scintillator, indicates that this may be feasible at experiments like \acro{borexino}, \acro{sno}+, and \acro{JUNO}. 
If directional capabilities are established for these detectors, daily and annual modulation of multiply scattering dark matter events can also be used to validate putative signals. 

\appendix

\section{Recasting bounds from ancient mica}
\label{app:mica}

In this appendix we recast data collected in Ref.~\cite{Price:1986ky} in a search for particle tracks in ancient muscovite mica (see \cite{Drukier:2018pdy} for a recent proposal to use other ancient minerals). 
No tracks were found, resulting in a bound on the flux of such particles,
\bea
\Phi_{\rm mica} \lesssim 10^{-18}~{\rm cm^{-2}~s^{-1}~sr^{-1}},
\eea 
which corresponds to a dark matter mass reach of $\mdm \lesssim 10^{26}~{\rm GeV}$. To leave a track in mica of density $\rho_{\rm mica}$ = 2.8~GeV/cm$^3$, transiting dark matter must deposit enough energy such that $\rho^{-1}_{\rm mica} dE/dx \gsim 2.4~{\rm GeV ~cm^2~g^{-1}}$, where $dE/dx \simeq n_{\rm O} \sigma_{ {\rm O} \chi} m_{\rm O} \vdm^2$ is the energy deposition rate of dark matter in mica for dark matter-oxygen scattering cross section $\sigma_{ {\rm O} \chi}$. 
Here we have conservatively assumed that mica is composed entirely of oxygen, and taken $m_{\chi} \gg m_{\rm O}$, the nuclear mass of oxygen. 
This determines a minimum dark matter-nucleon cross section to leave a detectable etched track in mica: 
\beq
\sigmaNDM^{\rm mica} \geq 6 \times 10^{-23}~{\rm cm^2}.
\eeq
 The energy deposition rate also determines the maximum dark matter-nucleon scattering cross section for which the dark matter, after passing through the earth overburden, transits mica quickly enough to etch a track :
\bea
\sigma_{\rm ob}^{\rm mica} \lesssim 2 \times 10^{-14}~{\rm cm^2}\left( \frac{m_\chi}{10^{21}~{\rm GeV}} \right)~.
\eea
Here we have used an overburden calculation similar to those detailed in Sec.~\ref{sec:ob}, and require that the dark matter can still etch a track in mica after traveling through 3 km of earth's crust.

\section*{Acknowledgments} 
 
For helpful discussions, we thank 
Mark Chen, 
Eugenio del Nobile,
Gopolang Mohlabeng,
and 
Alex Wright. 
We also thank Francesco Ronga for pointing out the potential sensitivity to \acro{MIMP}s of the liquid scintillator sub-detector of the \acro{MACRO} experiment.
Research at Perimeter Institute is supported by the Government of Canada through Industry Canada and by the Province of Ontario through the Ministry of Economic Development \& Innovation. 
J.~B.,~B.~B.,~and N.R.~acknowledge the support of the Natural Sciences and Engineering Research Council of Canada. 
J.~B.~thanks the Aspen Center for Physics, which is supported by National Science Foundation grant PHY-1066293.
J.~K.~is supported by Department of Energy grant DE-SC0010504.
R.F.L.~is supported through grant~PHY-1719271 from the National Science Foundation.
J.~B., J.~K.,~and N.~R.~thank the organizers of Santa Fe Summer Workshop 2018. 
T\acro{RIUMF} receives federal funding
via a contribution agreement with the National Research Council Canada.

\bibliography{nuxmas}

\end{document}